\documentclass[10pt,a4paper]{article}
\usepackage[left=10mm, right=10mm, top=20mm, bottom=20mm]{geometry}
\usepackage[utf8]{inputenc}
\usepackage{kbordermatrix}
\usepackage{amsmath}
\usepackage{amsfonts}
\usepackage{amssymb}
\usepackage{amsthm}
\usepackage{makeidx}
\usepackage{graphicx}
\usepackage{caption}
\usepackage{subcaption}
\usepackage{braket}
\usepackage{multirow}
\usepackage{tikz}
\usepackage{siunitx}
\usepackage{url}
\usepackage{float}
\usepackage{xcolor}
\usepackage{array}
\usepackage[sort,numbers]{natbib}
\usepackage{authblk}
\usepackage{natbib}

\renewcommand{\kbrdelim}{)}
\renewcommand{\kbldelim}{(}

\title{All magnetic field values cancelling $D_1$ line transitions of alkali metal atoms}
\author[*,1,2]{Artur Aleksanyan}
\author[1,2]{Rodolphe Momier}
\author[1,3]{Emil Gazazyan}
\author[1]{Aram Papoyan}
\author[2]{Claude Leroy}
\affil[1]{Institute for Physical Research, NAS of Armenia, Ashtarak-2, 0203, Armenia}
\affil[2]{Laboratoire Interdisciplinaire Carnot de Bourgogne, UMR CNRS 6303, Universit\'e de Bourgogne Franche-Comt\'e, 21000 Dijon, France}
\affil[3]{Institute for Informatics and Automation Problems, NAS of Armenia, Yerevan, 0014, Armenia}
\affil[*]{Corresponding author: arthuraleksan@gmail.com}

\date{\today}
\date{\vspace{-5ex}}

\begin{document}
\maketitle

\begin{abstract}
 In this work, $\pi$, $\sigma^+$ and $\sigma^-$ transitions between magnetic sublevels of the $D_1$ line of all alkali atoms are considered analytically. General block Hamiltonian matrices in presence of a magnetic field for the ground and excited states are built in order to describe all the transitions. Eigenvalues and eigenkets describing ground and excited levels are calculated, ``modified" and unperturbed transfer coefficients as a function of the nuclear spin $I$, the magnetic quantum number $m$ and the magnetic field magnitude $B$ are defined. Transition cancellations are observed only for some $\pi$ transitions of each isotope. The main result is that we obtain one single formula which expresses the magnetic field values cancelling these transitions. These values also correspond to the case when some of other transitions intensity have their maximum. In addition, we examine the derivative of $\pi$ transition ``modified" transfer coefficients in order to find the magnetic field values which correspond to the intensities maximum. The accuracy of the magnetic field $B$ values is only limited by the uncertainty of the involved physical quantities.
\end{abstract}

\section{Introduction}
Alkali–metal vapors are widely used in atomic physics: in laser experiments \cite{Sargsyan2015}, information storage \cite{Legaie2018}, spectroscopy \cite{Arimondo,AuzinshBook,Aleksanyan2020}, magnetometry~\cite{Yin_2021,Guo}, laser frequency stabilization~\cite{Talker} and are also the main material to study Bose-Einstein condensates~\cite{Bertrand,Maurus}. All of this is due to the fact that alkali metal atoms have a high transition intensity close to the infrared range. Cw narrow–band diode lasers operating in this domain have good features and are cheap, which allow experimenters to make the experiments easier. The above-mentioned properties make the study of alkali metal vapor transitions very important, especially in an external magnetic field.

It is well known that in a moderate external magnetic field $B$, atomic energy levels split into magnetic sublevels (Zeeman splitting). The frequency difference between ground and excited sub–levels is greatly deviated from the linear behavior \cite{Tremblay,Papageorgiou1994,Sargsyan2008}. Significant changes also occur for atomic transition probabilities. In the case of small values of $B$ (up to $\approx$~1000~G), the Zeeman splitted hyperfine transitions are overlapped because of Doppler broadening. To study the behavior of each atomic transition one should use sub–Doppler techniques \cite{Khanbekyan2016,Sargsyan2019}. It was demonstrated \cite{Sargsyan2017} that by derivative selective reflection strong line narrowing can be achieved.

In this paper ``modified" and unperturbed transition coefficients for $D_1$ line are analytically obtained. General formula for the magnetic field values cancelling some transitions while leading to the maximum of others is extracted. We determine an expression describing cancellation or maximum value of the transitions for the same magnetic quantum number $m$.

From the point of view of numerical simulations, all stable and long-lived isotope ``modified" transfer coefficients within a magnetic field are examined and $B$ field values for which the transition intensities are maximum or zero are calculated.

\section{Theory}
Fine structure is the splitting of main spectral lines of an atom. It is the result of the coupling between the orbital angular momentum $\boldsymbol{L}$ and spin angular momentum $\boldsymbol{S}$ of the single optical electron. The total electron angular momentum can be written in the following form:
\begin{equation}
\boldsymbol{J}=\boldsymbol{L}+\boldsymbol{S}.
\end{equation}
For $s \rightarrow p$ transitions, for the state we call ground we have $L=0$ and $S=1/2$ and for the excited state $L=1$ and $S=1/2$.

The hyperfine structure is the result of the combination between the total electron angular momentum $\boldsymbol{J}$ and the total nuclear angular momentum $\boldsymbol{I}$ of the atom. The total angular momentum $\boldsymbol{F}$ is the vector sum of $\boldsymbol{I}$ and $\boldsymbol{J}$:
\begin{equation}
\boldsymbol{F}=\boldsymbol{I}+\boldsymbol{J}.
\end{equation}
In this work only the $D_1$ line of alkali atoms is considered. In these cases the value of the total electron angular momentum magnitude is $J=1/2$. The total atomic angular momentum magnitude takes the following values:
\begin{equation}
I-1/2 \leq F \leq I+1/2,
\end{equation}
where $I$ is the magnitude of the total nuclear angular momentum. For all alkali atoms the total nuclear angular momentum is an integer or half-integer quantity. For the $F$ number the following notations will be used:
\begin{equation}
\begin{split}
F_{g,e}^{\pm}=I\pm1/2,
\end{split}
\end{equation}
where indices $g$ and $e$ stand for the ground and excited states respectively.

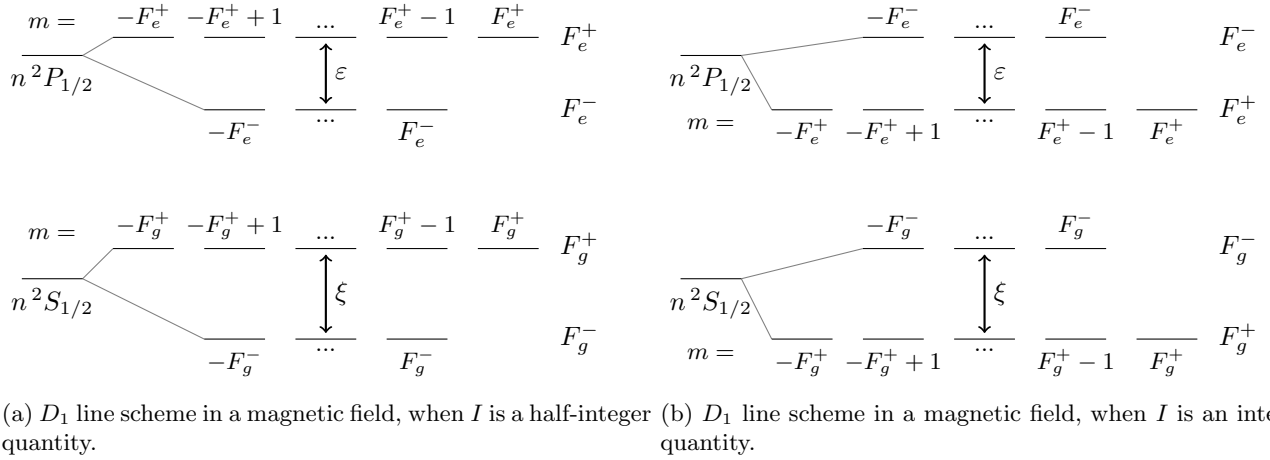
\begin{figure}[H]
\centering
\begin{subfigure}[b]{0.45\textwidth}
\begin{tikzpicture}[scale=0.8]
\draw[gray] (-0.5,4.7) -- (0,5); \draw[gray] (-0.5,4.7) -- (1.5,3.8); \draw[gray] (-0.5,1) -- (0,1.5); \draw[gray] (-0.5,1) -- (1.5,0);

\node [above] at (-1,5) {\small $m=$};
\draw (0,5) -- (1,5); \node [above] at (0.5,5) {\small $-F_e^+$};
\draw (1.5,5) -- (2.5,5); \node [above] at (2,5) {\small $-F_e^++1$};
\draw (3,5) -- (4,5); \node [above] at (3.5,5) {\small $...$};
\draw (4.5,5) -- (5.5,5); \node [above] at (5,5) {\small $F_e^+-1$};
\draw (6,5) -- (7,5); \node [above] at (6.5,5) {\small $F_e^+$};
\node [right] at (7.2,5) {$F_e^+$};

\draw (1.5,3.8) -- (2.5,3.8); \node [below] at (2,3.8) {\small $-F_e^-$}; \draw (3,3.8) -- (4,3.8); \node [below] at (3.5,3.8) {\small $...$};
\draw (4.5,3.8) -- (5.5,3.8); \node [below] at (5,3.8) {$F_e^-$};
\node [right] at (7.2,3.8) {$F_e^-$};

\node [above] at (-1,1.5) {\small $m=$};
\draw (0,1.5) -- (1,1.5); \node [above] at (0.5,1.5) {\small $-F_g^+$};
\draw (1.5,1.5) -- (2.5,1.5); \node [above] at (2,1.5) {\small $-F_g^++1$};
\draw (3,1.5) -- (4,1.5); \node [above] at (3.5,1.5) {\small $...$};
\draw (4.5,1.5) -- (5.5,1.5); \node [above] at (5,1.5) {\small $F_g^+-1$}; \draw (6,1.5) -- (7,1.5); \node [above] at (6.5,1.5) {\small $F_g^+$}; \node [right] at (7.2,1.5) {$F_g^+$};

\draw (1.5,0) -- (2.5,0); \node [below] at (2,0) {\small $-F_g^-$};
\draw (3,0) -- (4,0); \node [below] at (3.5,0) {\small $...$};
\draw (4.5,0) -- (5.5,0);; \node [below] at (5,0) {\small $F_g^-$}; \node [right] at (7.2,0) {$F_g^-$};

\draw [thick] [<->] (3.5,0.1) -- (3.5,1.4); \node [right] at (3.5,0.75) {$\xi$};
\draw [thick] [<->] (3.5,3.9) -- (3.5,4.9); \node [right] at (3.5,4.4) {$\varepsilon$};

\draw (-1.5,4.7) -- (-0.5,4.7); \node [below] at (-1,4.7) {$n\,^2P_{1/2}$};
\draw (-1.5,1) -- (-0.5,1); \node [below] at (-1,1) {$n\,^2S_{1/2}$};
\end{tikzpicture}
\caption[]{$D_1$ line scheme in a magnetic field, when $I$ is a half-integer quantity.}
\label{fig:I_half_D1_scheme}
\end{subfigure}
\begin{subfigure}[b]{0.45\textwidth}
\begin{tikzpicture}[scale=0.8]
\draw[gray] (-0.5,4.7) -- (0,3.8); \draw[gray] (-0.5,4.7) -- (1.5,5); \draw[gray] (-0.5,1) -- (0,0); \draw[gray] (-0.5,1) -- (1.5,1.5);

\node [below] at (-1,3.8) {\small $m=$};
\draw (1.5,5) -- (2.5,5); \node [above] at (2,5) {\small $-F_e^-$};
\draw (3,5) -- (4,5); \node [above] at (3.5,5) {\small $...$};
\draw (4.5,5) -- (5.5,5); \node [above] at (5,5) {\small $F_e^-$};
\node [right] at (7.2,5) {$F_e^-$};

\draw (0,3.8) -- (1,3.8); \node [below] at (0.5,3.8) {\small $-F_e^+$};
\draw (1.5,3.8) -- (2.5,3.8); \node [below] at (2,3.8) {\small $-F_e^++1$};
\draw (3,3.8) -- (4,3.8); \node [below] at (3.5,3.8) {\small $...$};
\draw (4.5,3.8) -- (5.5,3.8); \node [below] at (5,3.8) {\small $F_e^+-1$}; \draw (6,3.8) -- (7,3.8); \node [below] at (6.5,3.8) {\small $F_e^+$}; \node [right] at (7.2,3.8) {$F_e^+$};

\draw (1.5,1.5) -- (2.5,1.5); \node [above] at (2,1.5) {\small $-F_g^-$};
\draw (3,1.5) -- (4,1.5); \node [above] at (3.5,1.5) {\small $...$};
\draw (4.5,1.5) -- (5.5,1.5); \node [above] at (5,1.5) {\small $F_g^-$}; \node [right] at (7.2,1.5) {$F_g^-$};

\node [below] at (-1,0) {\small $m=$};
\draw (0,0) -- (1,0); \node [below] at (0.5,0) {\small $-F_g^+$};
\draw (1.5,0) -- (2.5,0); \node [below] at (2,0) {\small $-F_g^++1$};
\draw (3,0) -- (4,0); \node [below] at (3.5,0) {\small $...$};
\draw (4.5,0) -- (5.5,0); \node [below] at (5,0) {\small $F_g^+-1$};
\draw (6,0) -- (7,0); \node [below] at (6.5,0) {\small $F_g^+$}; \node [right] at (7.2,0) {$F_g^+$};

\draw [thick] [<->] (3.5,0.1) -- (3.5,1.4); \node [right] at (3.5,0.75) {$\xi$};
\draw [thick] [<->] (3.5,3.9) -- (3.5,4.9); \node [right] at (3.5,4.4) {$\varepsilon$};

\draw (-1.5,4.7) -- (-0.5,4.7); \node [below] at (-1,4.7) {$n\,^2P_{1/2}$};
\draw (-1.5,1) -- (-0.5,1); \node [below] at (-1,1) {$n\,^2S_{1/2}$};
\end{tikzpicture}
\caption[]{$D_1$ line scheme in a magnetic field, when $I$ is an integer quantity.}
\label{fig:I_nat_D1_scheme}
\end{subfigure}
\caption[]{Schemes of all possible $D_1$ lines within a magnetic field.}
\label{fig:D1_schemes}
\end{figure}

Within a magnetic field, the $D_1$ line energy levels splits into several magnetic sublevels, which are described by magnetic quantum numbers $m$. $m$ can take the following values:
\begin{equation}
-F \leq m \leq F.
\label{eq:m_range}
\end{equation}
On Fig.~\ref{fig:D1_schemes}, all possible $D_1$ lines schemes are depicted. The following notations are used: $n$ is the principal quantum number, which generally describes the system, $\xi=E_0(F_g^+)-E(F_g^-)$ is the energy difference between ground levels and $\varepsilon=E_0(F_e^+)-E(F_e^-)$ is the energy difference between excited levels.
It should be noted that when $I$ is an integer number the hyperfine structure is inverted.

Within a static magnetic field $\boldsymbol{B}$, the Hamiltonian $\mathcal{H}$ is the sum of the unperturbed Hamiltonian and the Zeeman Hamiltonian. We choose the direction of quantization axis the same as the direction of magnetic field \cite{Tremblay}. Taking into account the value of $J$, in the unperturbed basis $\ket{F,m}$, the diagonal elements of the Hamiltonian matrix $\mathcal{H}$ are
\begin{equation}
\bra{F,m}\mathcal{H}\ket{F,m}= E_{0}(F)-\mu_B g_{F}(F) m B,
\label{eq:Hamiltonian_diagonal_elements}
\end{equation}
where $E_0(F)$ is the zero-field energy of the level with total angular momentum value $F$, $\mu_B$ is the Bohr magneton, $g_{F}(F)$ is the associated Land\'e factor and $B$ is the magnitude of the magnetic field. Non-diagonal elements can be expressed in the following form:
\begin{equation}
\bra{F,m}\mathcal{H}\ket{F-1,m}=\bra{F-1,m}\mathcal{H}\ket{F,m}=-\frac{\mu_B}{2}\left(g_{J}-g_{I}\right) B \sqrt{1-\left(\dfrac{2m}{1+2I}\right)^2},
\label{eq:Hamiltonian_non_diagonal_elements}
\end{equation}
where $g_J$ and $g_I$ are the total angular and nuclear Land\'e factors \cite{HansBook} respectively. For the ground and excited states
\begin{equation}
g_J^g=g_S \text{ \ and \ } g_J^e=\dfrac{4g_L-g_S}{3}.
\end{equation}
As $F$ quantum numbers for ground and excited states are the same, in Eq.~\eqref{eq:Hamiltonian_diagonal_elements} we can use the following formulas for $g_F(F)$:
\begin{equation}
g_{F}(F_{g,e}^-)=g_{I}+ \dfrac{g_I-g_J^{g,e}}{1+2I} \text{ \ and \ } g_{F}(F_{g,e}^+)=\dfrac{g_J^{g,e}+2g_I I}{1+2I}.
\end{equation}
As we consider $D_1$ line transitions within a magnetic field, for a complete description of the system it is enough to write general $2 \times 2$ block matrices for the ground and excited states, where each block matrix corresponds to a given value of $m$, using Eq.~\eqref{eq:Hamiltonian_diagonal_elements} and Eq.~\eqref{eq:Hamiltonian_non_diagonal_elements}. We will not write Hamiltonian elements for the $m=\pm F_{g,e}$ values because they correspond to pure states and the corresponding transitions do not depend on magnetic field value $B$. The zero-field energies $E_0$ have been put to zero for both ground and excited states lower levels. Below, the matrix $\mathcal{H}_G$ describes the ground state and can be written as follows:
\begin{equation}
\mathcal{H}_G=
\kbordermatrix{\mbox{} & \ket{F^+_g,m_g} & \ket{F^-_g,m_g} \\
\bra{F^+_g,m_g} & \xi-\mu_B \dfrac{g_{S}+2 g_I I}{1+2I} m_g B
& \dfrac{\mu_B}{2}\left(g_{I}-g_{S}\right) B \sqrt{1-\left(\dfrac{2m_g}{1+2I}\right)^2} \\
\bra{F^-_g,m_g} & \dfrac{\mu_B}{2}\left(g_{I}-g_{S}\right) B \sqrt{1-\left(\dfrac{2m_g}{1+2I}\right)^2}
& -\mu_B \left(g_I+\dfrac{g_I-g_S}{1+2I}\right) m_g B \\
}.
\label{eq:G_matrix}
\end{equation}
For the excited state the general $2 \times 2$ Hamiltonian block matrix is 
\begin{equation}
\mathcal{H}_E=
\kbordermatrix{\mbox{} & \ket{F^+_e,m_e} & \ket{F^-_e,m_e} \\
\bra{F^+_e,m_e} & \varepsilon-\mu_B \dfrac{4g_L-g_S+6g_I I}{3(1+2I)} m_e B 
& \dfrac{\mu_B}{2} \cdot \dfrac{3g_{I}-4g_{L}+g_{S}}{3} B \sqrt{1-\left(\dfrac{2m_e}{1+2I}\right)^2} \\
\bra{F^-_e,m_e} & \dfrac{\mu_B}{2} \cdot \dfrac{3g_{I}-4g_{L}+g_{S}}{3} B \sqrt{1-\left(\dfrac{2m_e}{1+2I}\right)^2}
& -\mu_B \left(g_I+\dfrac{3g_I-4g_L+g_S}{3(1+2I)}\right) m_e B \\
}.
\label{eq:E_matrix}
\end{equation}
After diagonalization, the eigenvalues of the $\mathcal{H}_G$ matrix are given by
\begin{equation}
\Lambda_G^\pm=\dfrac{\xi-2\mu_B g_{I} m_g B}{2}\\ \pm \dfrac{1}{2}\left(\xi^2 + \mu_B^2 (g_I-g_S)^2 B^2 + \dfrac{4\xi\mu_B (g_I-g_S) m_g B}{1+2I}\right)^{1/2}.
\end{equation}
The eigenkets corresponding to $\Lambda_G^\pm$, expressed in terms of unperturbed state vectors $\ket{\psi(F_{g}, m_{g})}=\sum_{F_{g}^{\prime}} c_{F_{g} F^{\prime}_{g}}\ket{F_{g}^{\prime}, m_{g}}$ are
\begin{equation}
\ket{\psi(F_g^\pm,m_{g})}=\dfrac{1}{\sqrt{1+\kappa_{g\pm}^2}}\ket{F_g^+,m_{g}}+\dfrac{\kappa_{g\pm}}{\sqrt{1+\kappa_{g\pm}^2}}\ket{F_g^-,m_{g}},
\label{eq:eigket_ground}
\end{equation}
where we denoted $\kappa_{g\pm}=\dfrac{2(1+2I)(\Lambda_G^\pm-\xi)+2\mu_B (g_{S}+2 g_I I) m_g B}{\mu_B\left(g_{I}-g_{S}\right) B \sqrt{(1+2I)^2-4m_g^2}}$.

For the excited state, the eigenvalues of $\mathcal{H}_E$ are
\begin{equation}
\Lambda_E^\pm=\dfrac{\varepsilon-2\mu_B g_{I} m_e B}{2}\\ \pm \dfrac{1}{2}\left(\varepsilon^2 + \mu_B^2 \left(\dfrac{3g_I-4g_L+g_S}{3}\right)^2 B^2 + \dfrac{4\varepsilon\mu_B (3g_I-4g_L+g_S) m_e B}{3(1+2I)}\right)^{1/2},
\end{equation}
and its eigenkets, written in terms of the unperturbed atomic state vectors $\left|\psi\left(F_e, m_{e}\right)\right\rangle=\sum_{F_{e}^{\prime}} c_{F_{e} F^{\prime}_{e}}\ket{F_{e}^{\prime}, m_{e}}$, are the following:
\begin{equation}
\ket{\psi(F_e^{\pm},m_{e})}=\dfrac{1}{\sqrt{1+\kappa_{e\pm}^2}}\ket{F_e^+,m_{e}}+\dfrac{\kappa_{e\pm}}{\sqrt{1+\kappa_{e\pm}^2}}\ket{F_e^-,m_{e}},
\label{eq:eigket_excited}
\end{equation}
where $\kappa_{e\pm}=\dfrac{6(1+2I)(\Lambda_E^\pm-\varepsilon)+2\mu_B (4g_L-g_{S}+6 g_I I) m_e B}{\mu_B(3g_{I}-4g_L+g_{S}) B \sqrt{(1+2I)^2-4m_e^2}}$.

The relation which defines the electric dipole component $D_q$ \cite{Tremblay} is the following:
\begin{equation}
|\bra{e}D_{q}\ket{g}|^{2}=\frac{3 \varepsilon_{0} \hbar \Gamma_{e} \lambda_{e g}^{3}}{8 \pi^{2}} a^{2}[\ket{\psi(F_{e}, m_{e})} ; \ket{\psi(F_{g}, m_{g})} ; q],
\end{equation}
where $\varepsilon_{0}$ is the vacuum electric permittivity, $\Gamma_{e}$ is the natural decay rate, $\lambda_{e g}$ is the wavelength between ground and excited states, and $q=0,\pm 1$ corresponds to $\pi$, $\sigma^{\pm}$ transitions respectively. The definition of ``modified" transfer coefficient is
\begin{equation}
a[\ket{\psi(F_{e}, m_{e})} ; \ket{\psi(F_{g}, m_{g})} ; q]=\sum_{F_{e}^{\prime}, F_{g}^{\prime}} c_{F_{e} F_{e}^{\prime}} a(F_{e}^{\prime}, m_{e} ; F_{g}^{\prime}, m_{g} ; q) c_{F_{g} F_{g}^{\prime}},
\label{eq:a_coeff_sum}
\end{equation} 
where $a(F_{e}, m_{e} ; F_{g}, m_{g} ; q)$ are the unperturbed transfer coefficients:
\begin{equation}
a(F_{e}, m_{e} ; F_{g}, m_{g} ; q) =(-1)^{3/2+I+F_{e}+F_{g}-m_{e}}\sqrt{2}\sqrt{2 F_{e}+1}\sqrt{2 F_{g}+1} \left(\begin{array}{ccc}
{F_{e}} & {1} & {F_{g}} \\
{-m_{e}} & {q} & {m_{g}}
\end{array}\right)\left\{\begin{array}{ccc}
{F_{e}} & {1} & {F_{g}} \\
{1/2} & {I} & {1/2}
\end{array}\right\},
\label{eq:a_coeff_3j_6j}
\end{equation}
which depend on Wigner 3-j and 6-j symbols.

$\sigma^+$ and $\sigma^-$ transitions are not cancelled for any value of the magnetic field. Cancellations occur only for $\pi$ transitions. Let's examine the unperturbed transfer coefficients $a(F_{e}, m ; F_{g}, m ; 0)$. Two of them have the following expression:
\begin{equation}
a(F_e^\pm, m ; F_g^\pm, m ; 0)=\pm\dfrac{1}{\sqrt{3}} \cdot \dfrac{2m}{1+2I}.
\label{eq:unpert_trans_coeff_pi_+-+-}
\end{equation}
For the next two unperturbed coefficients the expression is
\begin{equation}
a(F_e^\pm, m ; F_g^\mp, m ; 0)=\dfrac{1}{\sqrt{3}} \cdot \sqrt{1-\left(\dfrac{2m}{1+2I}\right)^2}.
\label{eq:unpert_trans_coeff_pi_+--+}
\end{equation}

From Eq.~\eqref{eq:a_coeff_sum}, Eq.~\eqref{eq:a_coeff_3j_6j} and formulas \eqref{eq:unpert_trans_coeff_pi_+-+-}, \eqref{eq:unpert_trans_coeff_pi_+--+}, ``modified" transfer coefficients which have a cancellation are

\begin{multline}
a[\ket{\psi(F_e^\pm,m)},\ket{\psi(F_g^\pm,m)},0]=\\\dfrac{\kappa_{e\pm}}{\sqrt{1+\kappa_{e\pm}^2}} \times a\left(F_e^-,m;F_g^-,m;0\right) \times
\dfrac{\kappa_{g\pm}}{\sqrt{1+\kappa_{g\pm}^2}} 
+\dfrac{\kappa_{e\pm}}{\sqrt{1+\kappa_{e\pm}^2}} \times a\left(F_e^-,m;F_g^+,m;0\right) \times \dfrac{1}{\sqrt{1+\kappa_{g\pm}^2}}+ \\ \dfrac{1}{\sqrt{1+\kappa_{e\pm}^2}} \times a\left(F_e^+,m;F_g^-,m;0\right) \times
\dfrac{\kappa_{g\pm}}{\sqrt{1+\kappa_{g\pm}^2}}
+\dfrac{1}{\sqrt{1+\kappa_{e\pm}^2}} \times a\left(F_e^+,m;F_g^+,m;0\right) \times \dfrac{1}{\sqrt{1+\kappa_{g\pm}^2}}.
\label{eq:B--_equation}
\end{multline}
The solutions of $a[\ket{\psi(F_e^\pm,m)},\ket{\psi(F_g^\pm,m)},0]=0$ are
\begin{equation}
B_\pm^\pm=-\dfrac{1}{\mu_B} \cdot \dfrac{2m}{1+2I} \cdot \dfrac{2\xi\varepsilon}{(g_I-g_S)\varepsilon + \dfrac{3 g_I-4 g_L+g_S}{3}\xi}.
\label{eq:B++--_equation_solution}
\end{equation}

The condition on the considered ``modified" transfer coefficients solutions which defines permissible values of magnetic quantum number $m$ depends on nuclear spin:
\begin{equation}
0 \leq (-1)^{2I}m \leq I-\dfrac{1}{2}.
\label{eq:m_conditions}
\end{equation}

From the formula \eqref{eq:B++--_equation_solution} one can notice that for isotopes having a half-integer nuclear spin, transition cancellations exist for $\pi$ transitions between levels for which the magnetic quantum number is zero ($m=0$). But as the atomic states are degenerated, it is not possible to observe the cancellations of these transitions.

``Modified" transfer coefficients $a[\ket{\psi(F_e^\mp,m)},\ket{\psi(F_g^\pm,m)},0]$ can not be equal to zero, but these quantities have a very interesting behaviour. While for certain values of $B$, transition intensities corresponding to  $a[\ket{\psi(F_e^\pm,m)},\ket{\psi(F_g^\pm,m)},0]$ are zero (transition cancellation), the transition intensities corresponding to $a[\ket{\psi(F_e^\mp,m)},\ket{\psi(F_g^\pm,m)},0]$ reach their maximum value which corresponds to the intensity of a $\pi$ transition occurring between pure states (so-called ``guiding" atomic transitions~\cite{Sargsyan2015AlkaliMA}). This is ensured by the calculation of the derivative of ``modified" transfer coefficients squared with respect to magnetic field:

\begin{equation}
\dfrac{\mathrm{d} a^2[\ket{\psi(F_e^\mp,m)},\ket{\psi(F_g^\pm,m)},0]}{\mathrm{d} B}=0.
\label{eq:derivative_of_a}
\end{equation}
The solution of Eq.~\eqref{eq:derivative_of_a} is exactly Eq.~\eqref{eq:B++--_equation_solution} with the condition mentioned in Eq.~\eqref{eq:m_conditions}.

We call quantities $a[\ket{\psi(F_e^\mp,m)},\ket{\psi(F_g^\pm,m)}]$ and $a[\ket{\psi(F_e^\pm,m)},\ket{\psi(F_g^\pm,m)}]$ pair-``unperturbed" transfer coefficients, and transitions corresponding to them pair-transitions. As one can notice, cancellations occur only for transitions obeying $\Delta F = F_e - F_g = 0$ and maximum values take place when $\Delta F = F_e - F_g = \pm 1$.

\section{Stable and long-lived isotopes analysis}
In this section we fully analyze $D_1$ line transition cancellations and maxima of $^{23}$Na, $^{39}$K, $^{40}$K, $^{41}$K, $^{85}$Rb, $^{87}$Rb and $^{133}$Cs atoms. All the mentioned isotopes except $^{40}$K and $^{87}$Rb are stable. The half-life of $^{40}$K is 1.248(3) and that of $^{87}$Rb is 49.23(22) billion years. It should be noted that we do not study all the possible isotopes of all alkali atoms mainly due to the lack of data on these isotopes and also because their half-life time is too short to envisage an experiment in the close future. However the present theory is still valid to study them.

In Table \ref{tab:isotopes_data} all considered isotope data is given with uncertainties. As one can see, the most imprecise values in general are $\varepsilon$. But for $^{39}$K, $^{40}$K and $^{41}$K frequency differences between ground state levels are not precisely known. These quantities have the most influence on the size of the uncertainties of the calculated $B$ values.
\begin{table}[H]
\caption[]{Values used to calculate transitions between $D_1$ line magnetic sublevels with their uncertainties.}
\begin{center}
\begin{tabular}{cccccc}
\hline
Isotope & $I$ & $g_L$ & $g_I$ \cite{Arimondo} & $\xi$ (MHz) & $\varepsilon$ (MHz)\\
\hline
$^{23}$Na & 3/2 & 0.99997613 \cite{SteckNa} & -0.00080461080(80) & 1771.6261288(10) \cite{Arimondo} & 188.697(14) \cite{Das2008, Wijngaarden1994}\\\hline
$^{39}$K & 3/2 & 0.99997905339670(14)* & -0.00014193489(12) & 461.73(14) \cite{Falke2006} & 57.696(10) \cite{Das2008}\\\hline
$^{40}$K & 4 & 0.99997974531640(14)* & 0.000176490(34) & -1285.87(35) \cite{Falke2006} & -155.31(35) \cite{Falke2006}\\\hline
$^{41}$K & 3/2 & 0.99998039390246(13)* & -0.00007790600(8) & 253.99(12) \cite{Falke2006, Arimondo, Tiecke} & 30.50(16) \cite{Falke2006} \\\hline
$^{85}$Rb & 5/2 & 0.99999354 \cite{SteckRb85} & -0.00029364000(60) & 3035.7324390(60) \cite{Arimondo} & 361.58(17) \cite{Banerjee, Barwood1991} \\\hline
$^{87}$Rb & 3/2 & 0.99999369 \cite{SteckRb87} & -0.0009951414(10) & 6834.682610904290(90) \cite{Bize_1999} & 814.50(13) \cite{Banerjee,Barwood1991,Arimondo} \\\hline
$^{133}$Cs & 7/2 & 0.99999587 \cite{SteckCs} & -0.00039885395(52) & 9192.631770 (exact) \cite{SteckCs} & 1167.680(30) \cite{Udem1999, Rafac1997} \\\hline
\end{tabular}
\end{center}
\label{tab:isotopes_data}
\end{table}
It should be noted that when the value of $I$ is integer (only for $^{40}$K in this work), the values of $\xi$ and $\varepsilon$ should have the minus sign to be in agreement with our notations. For further calculations, for the Bohr magneton and $g_S$ spin Land\'e factor we used $\mu_B/h=-1.3996245042(86)$~MHz/G and $g_S=2.0023193043622(15)$ \cite{MohrCODATA2014} values respectively.
In Table \ref{tab:isotopes_data}, * stands for the calculated values of $g_L$ using the exact formula of Phillips \cite{Phillips1949} and values for the isotopes of Audi \textit{et al.} \cite{Audi2003}. We noticed that in the paper of Phillips \cite{Phillips1949} $1/m$ is missing in the second term of the exact formula for $g_L$.

\begin{figure}[H]
    \centering
    \includegraphics[scale=0.8]{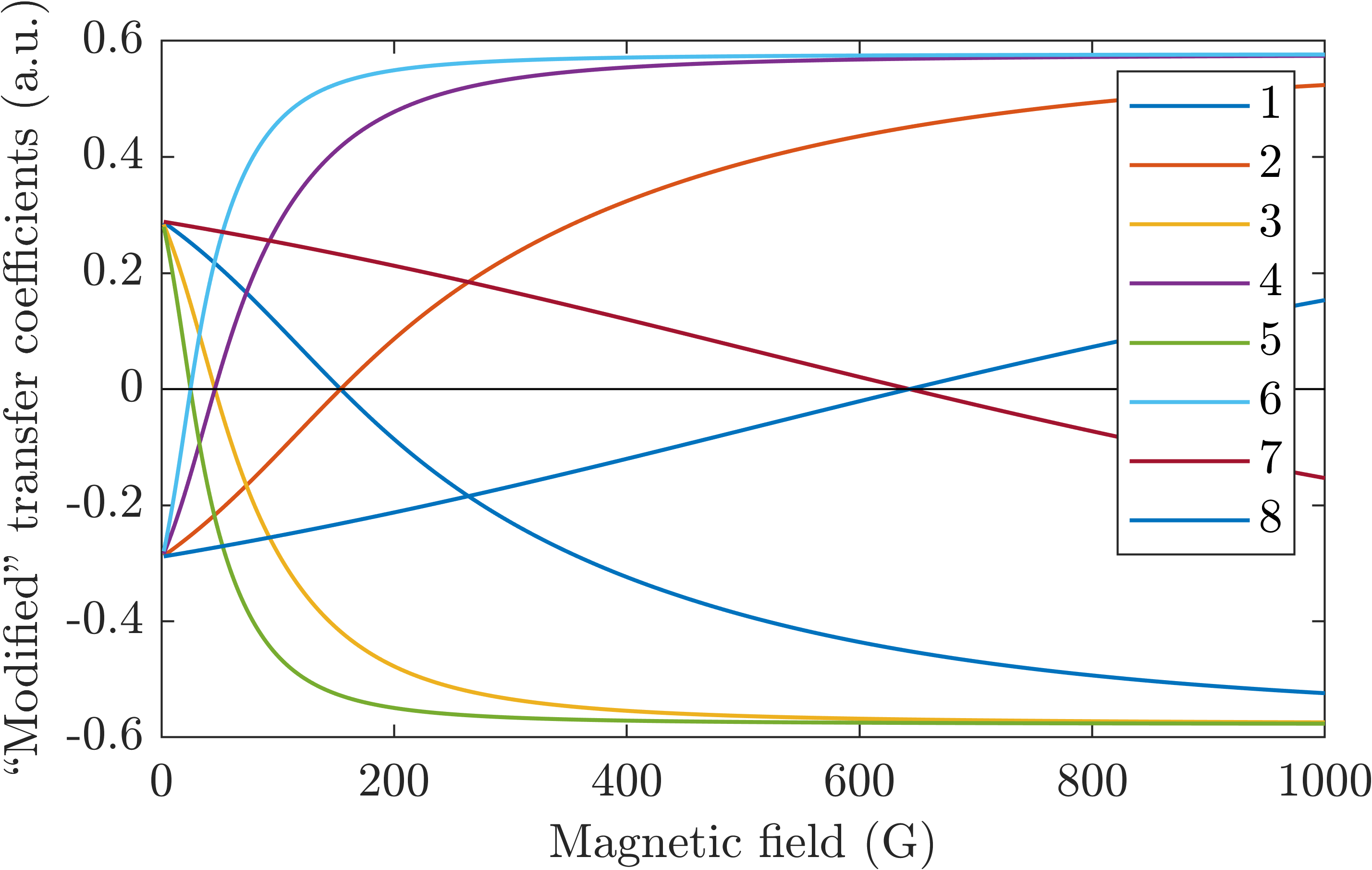}
    \caption{$^{23}$Na, $^{39}$K, $^{41}$K and $^{87}$Rb isotopes $D_1$ line $\pi$ transitions ``modified" transfer coefficients for $m=-1$. Numbering given in the inset is defined in Table~\ref{tab:cancellation_23Na_39K_41K_87Rb}.}
     \label{fig:graph_23Na_39K_41K_87Rb}
\end{figure}

For $^{23}$Na, $^{39}$K, $^{41}$K and $^{87}$Rb the total atomic angular momentum magnitude is $F=1$ for the lower levels of ground and excited states and $F=2$ for the upper levels. For all these isotopes, transitions cancel only for $m=-1$ (see Fig.~\ref{fig:graph_23Na_39K_41K_87Rb}).
\begin{table}[H]
\caption[]{$B$ field values cancelling transitions of $^{23}$Na, $^{39}$K, $^{41}$K and $^{87}$Rb with their uncertainties.}
\begin{center}
\begin{tabular}{cccccc}
\hline
Isotope & No. & $F$ & $m$ & $B$ (G)\\
\hline
$^{23}$Na & \begin{tabular}{@{}c@{}} 1\\2\end{tabular} & \begin{tabular}{@{}c@{}} 1\\2\end{tabular} & \begin{tabular}{@{}c@{}} -1\\-1\end{tabular} & \begin{tabular}{@{}c@{}} 153.2007(86)\\153.2007(86)\end{tabular} \\\hline
$^{39}$K & \begin{tabular}{@{}c@{}} 3\\4\end{tabular} & \begin{tabular}{@{}c@{}} 1\\2\end{tabular} & \begin{tabular}{@{}c@{}} -1\\-1\end{tabular} & \begin{tabular}{@{}c@{}} 44.991(10)\\44.991(10) \end{tabular} \\\hline
$^{41}$K & \begin{tabular}{@{}c@{}} 5\\6\end{tabular} & \begin{tabular}{@{}c@{}} 1\\2\end{tabular} & \begin{tabular}{@{}c@{}} -1\\-1\end{tabular} & \begin{tabular}{@{}c@{}} 24.046(95)\\24.046(95)\end{tabular} \\\hline
$^{87}$Rb & \begin{tabular}{@{}c@{}} 7\\8\end{tabular} & \begin{tabular}{@{}c@{}} 1\\2\end{tabular} & \begin{tabular}{@{}c@{}} -1\\-1\end{tabular} & \begin{tabular}{@{}c@{}} 642.590(76)\\642.590(76)\end{tabular} \\\hline
\end{tabular}
\end{center}
\label{tab:cancellation_23Na_39K_41K_87Rb}
\end{table}
In Table \ref{tab:cancellation_23Na_39K_41K_87Rb} all $B$ field values which cancel $D_1$ line transitions of $^{23}$Na, $^{39}$K, $^{41}$K and $^{87}$Rb are calculated. The numbers in the second column refer to the labeling of Fig.~\ref{fig:graph_23Na_39K_41K_87Rb}, the third column shows the values of the total angular momentum magnitude for both ground and excited states. The fourth column indicates between which magnetic sublevels the transition occurs and the fifth column displays calculated values of $B$ field with their uncertainties, which are a consequence of the physical quantities uncertainties involved in the calculations.

\begin{figure}[H]
    \centering
    \includegraphics[scale=0.1]{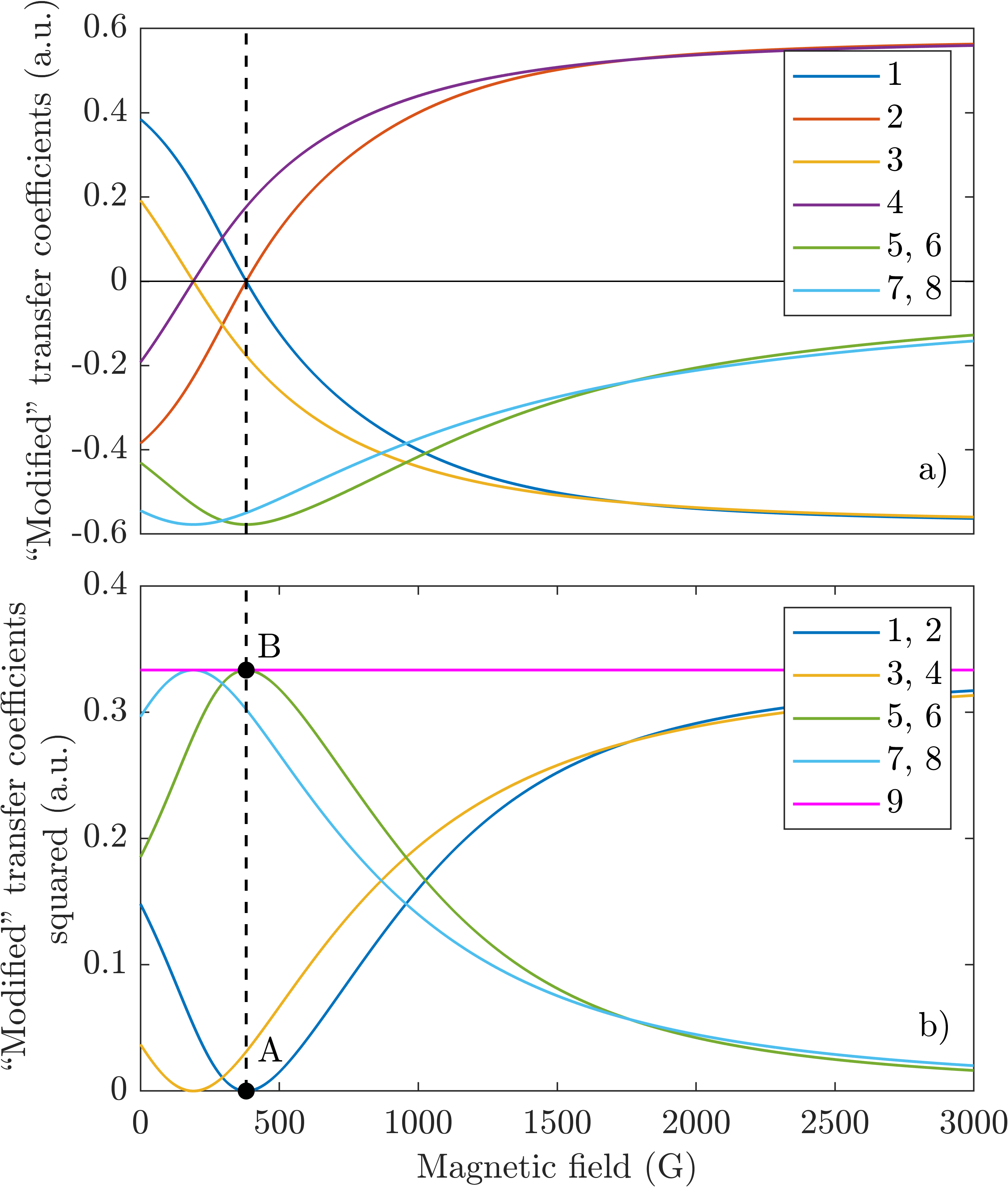}
    \caption{a) $^{85}$Rb $D_1$ line ``modified" transfer coefficients for $m=-2$ and $m=-1$ $\pi$ transitions. For this isotope four cancellations exist: 1, 2, 3 and 4. b) ``Modified" transfer coefficients squared for $m=-3,-2$ and $-1$. The vertical dashed line indicates the value $B=380.73$~G which corresponds to the cancellation of the transitions 1, 2 (point A) and coincides with the maximum of transitions 5, 6 (point B). Numbering 1-8 given in the inset is defined in~Table~\ref{tab:cancellation_85Rb}, ``guiding" transition coefficient squared 9 corresponds to $m=-3$.}
    \label{fig:graph_Rb_85}
\end{figure}

For $^{85}$Rb the total atomic angular momentum magnitude is $F=2$ for the lower levels of ground and excited states and $F=3$ for the upper levels. Some transitions cancel for $m=-2$ and $m=-1$. For this isotope we will also analyse those transitions which have maximum value as it was mentioned before. $\pi$ transitions corresponding to the above-mentioned magnetic quantum numbers which do not cancel reach their maximum values for the magnetic field values canceling the other transitions. On Fig.~\ref{fig:graph_Rb_85} a), ``modified" transfer coefficients (i.e. $a[\ket{\psi(F_e,m)},\ket{\psi(F_g,m)},0]$ quantities) are depicted for $m=-2$ and $m=-1$. Lines numbered 5, 6, 7 and 8 have no cancellation and are nothing more than transition coefficients between ground and excited states, where $F_g \neq F_e$. On Fig.~\ref{fig:graph_Rb_85} b), ``modified" transfer coefficients squared are depicted in order to compare them with each other and with ``guiding" atomic transition coefficient for which $m=-3$.

In Table \ref{tab:cancellation_85Rb} all $B$ field values which cancel transitions of $^{85}$Rb $D_1$ line are calculated. One also can see that for these values of magnetic field the transitions 5, 6, 7 and 8 reach their maximum value equal to the transfer coefficient 9 squared.

\begin{table}[H]
\caption[]{$B$ field values cancelling (1-4) or maximizing (5-8) transitions of $^{85}$Rb with their uncertainties.}
\begin{center}
\begin{tabular}{cc|ccc|ccc}
\hline
Isotope & $B$ (G) & No. & $F$ & $m$ & No. & $\Delta F$ & m\\
\hline
$^{85}$Rb & \begin{tabular}{@{}c@{}} 380.73(13)\\380.73(13)\\190.368(66)\\190.368(66) \end{tabular} & \begin{tabular}{@{}c@{}} 1\\2\\3\\4\end{tabular} & \begin{tabular}{@{}c@{}} 2\\3\\2\\3\end{tabular} & \begin{tabular}{@{}c@{}} -2\\-2\\-1\\-1\end{tabular} & \begin{tabular}{@{}c@{}} 5\\6\\7\\8 \end{tabular} & \begin{tabular}{@{}c@{}} -1\\1\\-1\\1\end{tabular} & \begin{tabular}{@{}c@{}} -2\\-2\\-1\\-1\end{tabular} \\\hline
\end{tabular}
\end{center}
\label{tab:cancellation_85Rb}
\end{table}

\begin{figure}[H]
    \centering
    \includegraphics[scale=0.8]{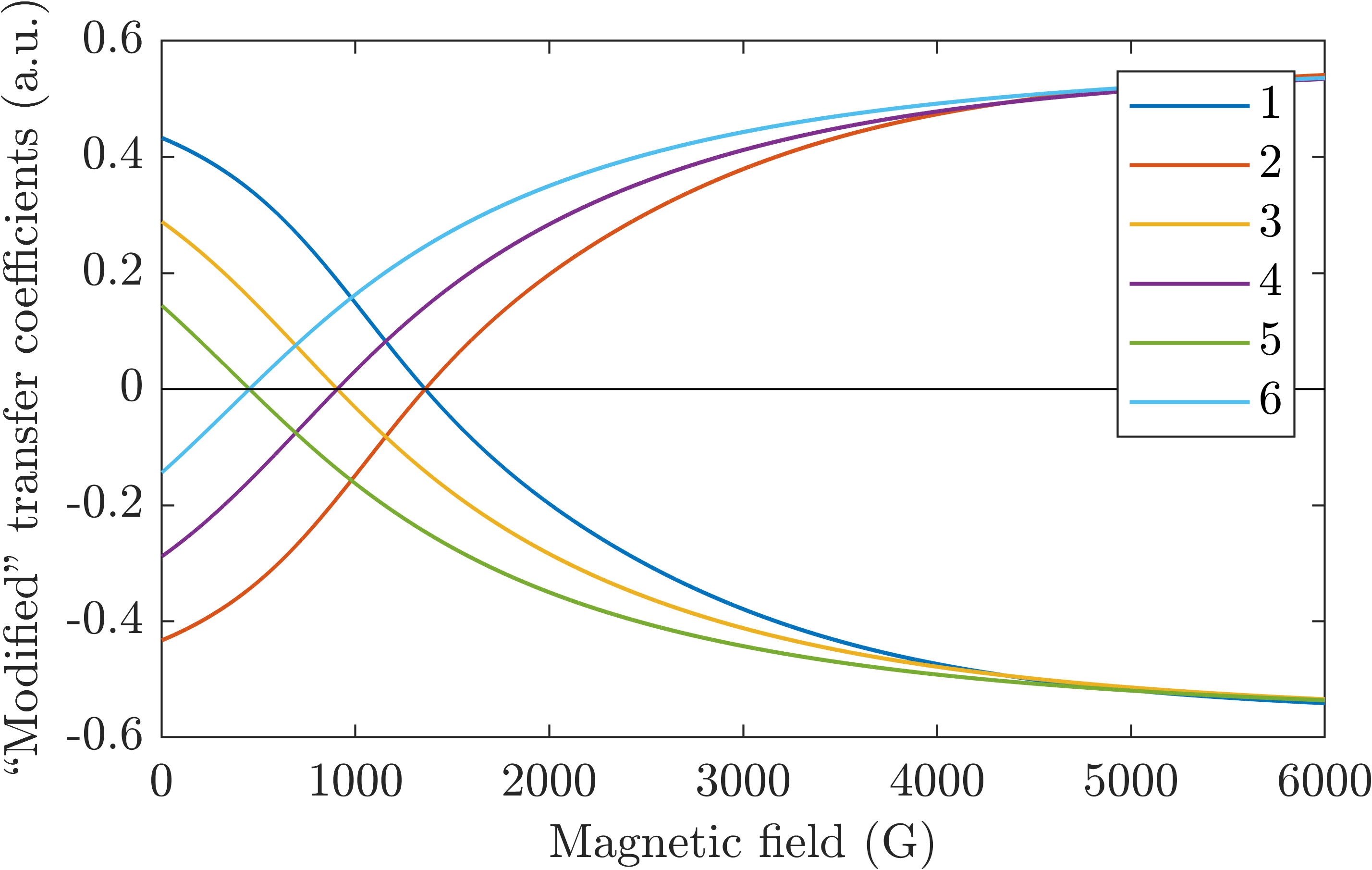}
    \caption{$^{133}$Cs $D_1$ line $\pi$ transition coefficients which have a cancellation. Cancellations are observed for $m=-3$, $m=-2$ and $m=-1$.}
    \label{fig:graph_Cs_133}
\end{figure}

\begin{figure}[H]
    \centering
    \includegraphics[scale=0.8]{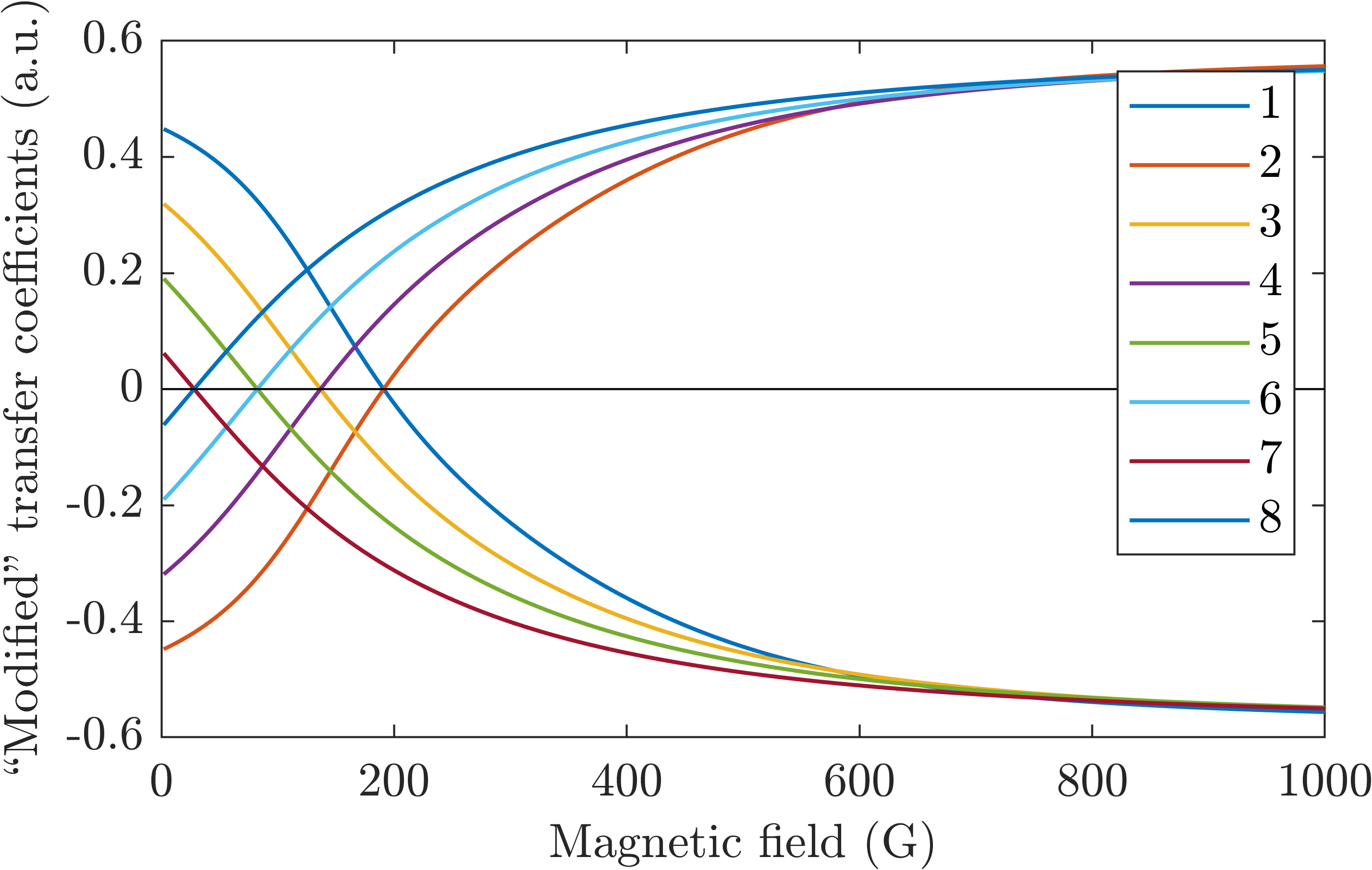}
    \caption{$^{40}$K $D_1$ line $\pi$ transition coefficients which have a cancellation. Cancellations are observed for $m=7/2$, $m=5/2$, $m=3/2$ and $m=1/2$.}
    \label{fig:graph_K_40}
\end{figure}

For $^{133}$Cs, the total atomic angular momentum magnitude is $F=3$ for the lower levels of ground and excited states and $F=4$ for the upper levels. Transition cancellations are observed for $m=-3$, $m=-2$ and $m=-1$. On Fig.~\ref{fig:graph_Cs_133}, ``modified" transfer coefficients for all $\pi$ transitions which have a cancellation are shown. 
In Table \ref{tab:cancellation_133Cs} all $B$ field values which cancel transitions of $^{133}$Cs $D_1$ line are calculated.
\begin{table}[H]
\caption[]{$B$ field values cancelling transitions of $^{133}$Cs with their uncertainties.}
\begin{center}
\begin{tabular}{cccccc}
\hline
Isotope & No. & $F$ & $m$ & $B$ (G)\\
\hline
$^{133}$Cs & \begin{tabular}{@{}c@{}} 1\\2\\3\\4\\5\\6\end{tabular} & \begin{tabular}{@{}c@{}} 3\\4\\3\\4\\3\\4\end{tabular} & \begin{tabular}{@{}c@{}} -3\\-3\\-2\\-2\\-1\\-1\end{tabular} & \begin{tabular}{@{}c@{}} 1359.237(26)\\1359.237(26)\\906.158(17)\\906.158(17)\\453.0790(84)\\453.0790(84) \end{tabular} \\\hline
\end{tabular}
\end{center}
\label{tab:cancellation_133Cs}
\end{table}
For $^{40}$K the total atomic angular momentum magnitude is $F=9/2$ for the lower levels of ground and excited states and $F=7/2$ for the upper levels. Transition cancellations are observed for $m=7/2$, $m=5/2$, $m=3/2$ and $m=1/2$. On Fig.~\ref{fig:graph_K_40}, ``modified" transfer coefficients for all $\pi$ transitions which have a cancellation are depicted.
In Table \ref{tab:cancellation_40K} all $B$ field values which cancel transitions of $^{40}$K $D_1$ line are calculated.
\begin{table}[H]
\caption[]{$B$ field values cancelling transitions of $^{40}$K with their uncertainties.}
\begin{center}
\begin{tabular}{cccccc}
\hline
Isotope & No. & $F$ & $m$ & $B$ (G)\\
\hline
$^{40}$K & \begin{tabular}{@{}c@{}} 1\\2\\3\\4\\5\\6\\7\\8\end{tabular} & \begin{tabular}{@{}c@{}} 9/2\\7/2\\9/2\\7/2\\9/2\\7/2\\9/2\\7/2\end{tabular} & \begin{tabular}{@{}c@{}} 7/2\\7/2\\5/2\\5/2\\3/2\\3/2\\1/2\\1/2\end{tabular} & \begin{tabular}{@{}c@{}} 190.20(33)\\190.20(33)\\135.85(24)\\135.85(24)\\81.51(15)\\81.51(15)\\27.171(48)\\27.171(48) \end{tabular} \\\hline
\end{tabular}
\end{center}
\label{tab:cancellation_40K}
\end{table}

\section{Conclusion and outlook}
In this work an analytical model to calculate all optical transitions within magnetic a field for all types of polarized light and for the $D_1$ line of all alkali atoms is used. We determined a unique formula expressing magnetic field values for which some $\pi$ transition intensities become zero and some others' intensities become maximum simultaneously. No $\sigma^+$ or $\sigma^-$ transitions have a cancellation. Several reasons lead us to find these values as precise as possible. The first one is that for very sensitive magnetometers calibration some standards should exist and these values are good standards \cite{Aleksanyan2020-1, Momier2020} for atomic systems: they do not depend on any external condition or parameter. The second reason is trying to improve the most imprecise parameter involved in our calculation: the energy difference between excited states. However, in the experiments it is always more complicated to precisely measure small signals than big ones, thus the cancellation of transitions can not be measured directly because of the existence of noise in any experiment. In other words, as for small peaks the signal-to-noise ratio is smaller than for higher peaks, it is profitable to measure peaks with bigger intensity. Thus in the experiment described in one of our latest papers \cite{Aleksanyan2020-1} we are going to measure those transitions (e.g. find maximums) which have a maximum value. If we will be able to find magnetic field magnitude for which the transition intensity is maximum, it will mean that we found pair-transition cancellation value. Obviously, the graph of the derivative of the intensity with respect to $B$ should be calculated in the neighbourhood of the maximum value, despite the fact that the change of transition intensity can be very smooth, as the change of sign of the slope of the derivative will give precisely the value for which it crosses the $B$-axis, thus will give the $B$ value for which the pair-transitions reach their minimum.

\noindent\textbf{Funding.} This research was sponsored in part by the NATO Science for Peace and Security Programme under grant G5794. Artur Aleksanyan acknowledges the funding support CO.17049.PAC.AN from the Graduate School EUR EIPHI.

\bibliographystyle{ieeetr}
\bibliography{references}

\end{document}